\documentclass[prd,superscriptaddress,twocolumn,preprintnumbers,%
  amsmath,amssymb]{revtex4-1}
  
\usepackage[breaklinks,colorlinks=true]{hyperref}
\usepackage{mathtools}
\usepackage{color}
\allowdisplaybreaks

\newcommand{\tr}{\mathrm{tr}\,}

\begin{document}
\title{Schwinger Pair Production in 
SL{\boldmath$(2,\mathbb{C})$} Topologically 
Non-Trivial Fields via Non-Abelian Worldline Instantons}
\author{Patrick Copinger}
\affiliation{Yukawa Institute for Theoretical Physics, Kyoto University, 
606-8502, Japan}
\email{patrick.copinger@yukawa.kyoto-u.ac.jp}
\author{Pablo Morales}
\affiliation{Research Division, Araya Inc., 
107-6019, Japan}
\email{pablo$_$morales@araya.org}

\preprint{YITP-20-150}

\begin{abstract}
Schwinger pair production is analyzed in a BPST instanton 
background and in its SL$(2,\mathbb{C})$ complex extension for 
complex scalar particles. A non-Abelian extension of the worldline 
instanton method is utilized, wherein Wong's equations in a coherent 
state picture adopted for SL$(2,\mathbb{C})$ are solved in Euclidean 
spacetime. While pair production is not predicted in the BPST 
instanton, a complex extension of the BPST instanton, existing as 
parallel fields in Minkowski spacetime, is shown to decay via the 
Schwinger effect.
\end{abstract}

\maketitle

\section{Introduction}

The quantum field theoretic (QFT) vacuum in a strong electric field
is thought unstable against the production of particle anti-particle 
pairs in what is known as the Schwinger 
mechanism~\cite{sauter,*Heisenberg:1935qt,*PhysRev.82.664}.
Observation of the Schwinger effect could be impactful for the 
understanding of non-perturbative QFTs, and is actively being or to be 
sought not only in strong quantum electrodynamics (QED) in 
high-power laser facilities (e.g., ELI-Beamlines, etc. reviewed 
in~\cite{danson_2019}), but also
in analog condensed matter settings, whereby the effect is facilitated 
through Landau-Zener transitions~\cite{doi:10.1098/rspa.1934.0116,
*10011873546,*doi:10.1098/rspa.1932.0165}. 
Yet, due to a strong exponential suppression, 
(i.e., $m^2c^3\pi/eE\hbar$ for 
homogeneous electric field, $E$, and scalar/fermion mass, $m$), 
the effect still has not been seen. While smallish in
high-power lasers, the Schwinger mechanism is thought to be 
a prominent feature of non-Abelian chromoelectric flux tube breaking in 
heavy-ion collisions~\cite{PhysRevD.20.179},
during which, in topologically non-trivial fields, chiral transport 
phenomena can develop.

The Schwinger mechanism has been argued to underlie the chiral 
anomaly for finite fermion mass 
systems~\cite{PhysRevLett.104.212001,*PhysRevD.86.085029}, and has 
been confirmed numerically~\cite{Tanji:2010eu,
*Andres:2018ifx,*Tanji:2018qws}. 
The axial Ward 
identity at operator level in QED reads 
$\partial_\mu j_5^\mu = -(e^2/16\pi^2) \epsilon^{\mu\nu\alpha\beta}
 F_{\mu\nu}F_{\alpha\beta} + 2m\bar{\psi} i \gamma_5\psi$, 
 with $j^\mu_5$ the axial current density and $\psi$ the fermion 
 field~\cite{PhysRev.177.2426,*ref1}. 
 Then in homogeneous fields with nonzero Chern-Pontryagin density, 
 the matrix element vacuum polarization, 
 (i.e., in-out expectation values), of 
 the axial Ward identity indicates an anomaly cancellation; however in-in 
 expectation values of the identity, properly accounting for the Schwinger 
 effect, restore the anomaly~\cite{PhysRevLett.121.261602,
 *doi:10.1142/S0217751X2030015X}. 
And, a pertinent and intuitive question one may ask is how does the 
anomaly behave under the Schwinger mechanism in non-Abelian and 
non-trivial topological fields?
To address this we must first understand the Schwinger effect for 
massive particles under such fields. The case of massless fermions 
under a isotropic and homogeneous SU$(2)$ gauge field background 
with a non-vanishing Chern-Pontryagin density leading to the chiral 
anomaly via the Schwinger effect was explored 
in~\cite{Domcke:2018gfr}. And massive pair production has also been 
studied in axion-SU$(2)$ field during 
inflation~\cite{PhysRevD.101.083528}.

The index theorem is well-known to relate the fermionic left and right 
zero modes to the Pontryagin number~\cite{ATIYAH1978185}. A key 
example~\cite{PhysRevD.14.3432} is provided through the Dirac operator 
in a Belavin, Polyakov, Schwarz and Tyupkin (BPST) 
instanton~\cite{Belavin197585,*RevModPhys.70.323,*Vandoren:2008xg}
background, and therefore the instanton is an intuitive choice of a 
topologically non-trivial background under which to study Schwinger pair 
production. A measure of pair 
production is provided through the vacuum non-persistence, namely the 
appearance of an imaginary part in the background field effective action; 
the non-persistence predicts the vacuum instability sum over any number 
of pair permutations~\cite{Fradkin:1991zq}. 
One may characterize the non-persistence as arising for the condition, 
$|\langle\Omega_{out} | \Omega_{in}\rangle|^2\neq 1$, for in and out 
asymptotic vacuum states. Then, as the BPST instanton interpolates 
between differing asymptotic winding numbers in Euclidean time, one 
should not dismiss out-of-hand a role played by the 
Schwinger mechanism. 
Nevertheless, an imaginary part of the fermion/boson determinant in a 
BPST instanton is not seen; they have been evaluated exactly 
in~\cite{PhysRevLett.94.072001,*PhysRevD.71.085019}. Not only do we 
seek to explain this, but moreover also to explore under what conditions 
does Schwinger pair production occur. 

Pair production under inhomogeneous background fields can be 
analytically cumbersome, but has been well studied using the 
non-perturbative worldline instanton (WI) 
method~\cite{AFFLECK1982509,PhysRevD.72.105004,
PhysRevD.73.065028}. Thus, it is instructive to study the more 
complicated topological fields through a non-Abelian extension of the WI 
method. To avoid confusion with the WIs, let us refer to disparate BPST 
instantons compactly as Yang Mills Instantons (YMI); anti-instantons are 
simply anti-YMI. The WI reminiscent Lorentz force equation is to an Abelian system what Wong's equations~\cite{Wong:1970fu,
PhysRevD.15.2308,PhysRevD.17.3247} are to a non-Abelian system.
And Wong's equations describe the classical evolution of a particle
in a Yang Mills field. To arrive at Wong's equations we employ a coherent 
state formalism~\cite{Zhang:1990fy} on the Wilson loop, converting the 
propertime ordered matrix weighted exponential into a path integral over 
the Haar measure. This process is known for the Non-Abelian Stokes 
Theorem (NAST)~\cite{PhysRevD.58.105016,
DIAKONOV1989131,*10.1143/PTP.104.1189,*KONDO20151} as well as for 
the chiral kinetic theory~\cite{PhysRevLett.109.162001}, with non-Abelian 
degrees of freedom~\cite{PhysRevD.88.045012,
*Dwivedi_2013,*PhysRevD.89.094003}.
The chiral kinetic theory has also been explored on the worldline~\cite{PhysRevD.96.016023,
PhysRevD.97.051901,PhysRevD.99.056003}. 
And worldline techniques under a coherent state formalism
have proved valuable for the calculation of polarized deeply inelastic 
scattering~\cite{PhysRevD.102.114022}.
Let us last point out that color degrees of freedom in the Wilson loop 
have also equivalently been described with auxilary fields; 
see~\cite{BARDUCCI1981141,*Bastianelli_2013,
*PhysRevD.93.025035,*PhysRevD.93.049904,*Bastianelli:2015iba} 
and~\cite{Corradini:2016czo,*Edwards:2016acz,*Ahmadiniaz:2018olx}
for the more general case.

What is novel in our approach is 
the extension of the coherent state formalism for the Wilson loop to a 
non-compact SL$(2,\mathbb{C})$ group.
There is a topological equivalence between the coherent
states and coset elements, here SL$(2,\mathbb{C})/$SU(2), that is 
manifest in the construction of the Hilbert space, $H$. 
Group complexifications of the type 
$H^{C}/H$--most remarkably SU$(2)$--have played a stellar role in 
various fields. For instance, precise formulation of the AdS/CFT 
correspondence requires Euclidean anti-de Sitter space AdS$_3$
string theory topologically equivalent
SL$(2,\mathbb{C})/$SU$(2)$~\cite{
Gawedzki:1991yu,*ISHIBASHI2000149,*TESCHNER1999390,
doi:10.1063/1.1377273,*TESCHNER1999369, *Satoh:1999jc}.
Also Chern-Simons gauge theory with complex gauge group
SL$(2, \mathbb{C})$~\cite{Gukov:2006ze}
has been found to exhibit many interesting connections with 
three-dimensional quantum gravity and the geometry of a hyperbolic 
three-manifold; see~\cite{Gukov:2003na} and references within.
Let us also comment that most spinfoam models for 4d gravity have been 
constructed as discretized path integrals for constrained background 
field theories with 
SL$(2,\mathbb{C})$~\cite{Dupuis:2010jn,*Dupuis:2011wy}.
In this paper we consider SU$(2)^{\mathbb{C}}$, 
extending the WI formalism to explore non-Abelian topologically 
non-trivial fields, and also through analytical continuation we parametrize 
the effective action.

The WI method has also been extended to finite 
temperature~\cite{PhysRevD.95.056006} and to worldline 
sphalerons~\cite{PhysRevD.96.076002,*PhysRevD.98.056022}. However,
we treat the effective action to one-loop at zero temperature, and we also 
negate backreaction effects. 
Last, we focus on complex scalar production.
In a YMI background the fermion functional determinant is proportional to 
that of a complex scalar~\cite{tHooft:1976snw,*PhysRevD.62.114022}, 
since the spectrums are similar apart from a multiplicity factor and zero 
modes. Furthermore, apart from a pre-factor the fermion effective action 
only differs from the complex scalar one through a spin factor, whose 
contributions can be safely neglected in performing the WI method for 
homogeneous, Sauter-type, or sinusoidal 
potentials~\cite{PhysRevD.72.105004}. 

This work is organized as follows: We begin with a cursory examination 
of the Schwinger effect, and its absence in SU$(2)$ Euclidean fields in 
Sec.~\ref{sec:mink_pp}. Then we develop the worldline formalism
for non-Abelian field in Sec.~\ref{sec:na_wi}. Then as a 
demonstration of the absence, we examine pair production in a YMI
in Sec~\ref{sec:pp_bpst}. And its complex extension, which does
yield pair production, is sought in Sec.~\ref{sec:pp_cbpst}. Conclusions 
are finally presented in Sec.~\ref{sec:conclusions}.

\section{Minkowski Electric Fields and the Schwinger Effect}
\label{sec:mink_pp}

Let us begin our discussion of pair production (or the lack thereof)
in a YMI by examining the large instanton limit, 
$R^2\gg x^2$ for instanton parameter $R$; this is for the field strength of 
the YMI in the regular gauge,
\begin{equation}
-\frac{4R^{2}}{g(x^{2}+R^{2})^{2}}\sigma_{\mu\nu}\rightarrow
-\frac{4}{gR^{2}}\sigma_{\mu\nu}\,;
\label{eq:large_instanton}
\end{equation}
we use conventional notations that are listed below. The large instanton 
limit was explored in~\cite{PhysRevD.85.045026,*Basar2013}. 
The key point here is that in the large instanton limit, the YMI resembles
a homogeneous field, one with (in a Minkowski spacetime picture) 
magnetic fields, 
$B_j=-(2/g)[R^{2}/(\rho^{2}+R^{2})^{2}]\sigma_{j}$, 
and imaginary electric fields, 
$E_j=i(2/g)[R^{2}/(\rho^{2}+R^{2})^{2}]\sigma_{j}$. 
As is well understood
for the Abelian homogeneous case, there can be no pair production
in sole constant magnetic fields~\footnote{One cavet here is that 
a strong magnetic field may well decay into 
magnetic monopoles~\cite{AFFLECK198238},
but not (two-color) quarks or electrons and positrons.}. 
And the imaginary
electric fields too act as a magnetic field, giving rise to no
poles in Schwinger propertime.
Therefore, it is anticipated there should be no pair production in
the YMI.

It is simple to see why this should be the case.
Pair production is governed by the non-persistence criteria,
which goes as, for a complex scalar,
$|\langle \Omega_{out}|\Omega_{in}\rangle |^2 =
|\det(-\mathcal{D}^2+m^2)|$. The determinant can be written as
a product of eigenvalues of the scalar operator. However, in a Euclidean
metric since the scalar operator under a YMI background is Hermitian,
its squared operator eigenvalues must be semi-positive definite.
And one would not expect to see a complex phase emerging from the 
determinant. Put another way, one must have the sign problem
in order to see Schwinger pair production, 
that is the sign problem in a worldline path integral formalism in a 
Euclidean metric, which is unambiguous. 
After Wick rotation, real Minkowski electric fields
become imaginary in Euclidean spacetime, giving rise to the sign
problem in the worldline action.

Through similar reasoning, we would anticipate a similar pair production 
absence for a number of topological Yang Mills solutions in SU$(2)$, 
such as for Wu-Yang monopoles~\cite{WU1976365}, 
merons~\cite{DEALFARO1976163,*DEALFARO1977203}, 
among others~\cite{RevModPhys.51.461}.
Sphalerons~\cite{PhysRevD.30.2212,*PhysRevD.36.581}, 
however, exist at finite temperature, to which the above arguments may 
not apply. But, it is curious to ask to what topological objects may 
Schwinger pair production play a role. In light of the large instanton limit
an intuitive surmise would be objects with real electric fields in 
Minkowski spacetime. We explore such objects below, in a natural
extension to the YMI. However, a real SU$(2)$ background field in 
Minkowski spacetime after Wick rotation becomes an 
SL$(2,\mathbb{C})$ field in Euclidean 
spacetime~\cite{PhysRevD.12.3843,*PhysRevD.13.3233}.
We apply the WI method to study pair production, and to make the 
extension to the most general complex field we apply the coherent
state method to non-compact groups, doing so furthermore
furnishes us with a worldline action amenable to Wong's equations
and WIs. Even if one were to not employ a Euclidean metric the 
extension to a complexified group is expected, indeed even for the
Abelian case, complex WIs are important~\cite{PhysRevD.84.125023}.

\section{Non-Abelian Worldline Instantons}
\label{sec:na_wi}

To more fully explore the Schwinger effect in non-trivial topological
fields let us build on the WI 
method~\cite{AFFLECK1982509,
PhysRevD.72.105004,PhysRevD.73.065028}.
Our starting point is the one-loop effective action for a complex scalar 
particle with mass, $m$, in a non-Abelian Euclidean background field,
\begin{equation}
\Gamma[A]=-\frac{1}{2}\log\det(-\mathcal{D}^2+m^2)\,,
\label{eq:eff_action_def}
\end{equation}
where in the fundamental representation we have 
$\mathcal{D}_\mu=\partial_\mu-igA_\mu (x)$
with $A_\mu (x)=A_\mu^a (x)T^a$. $G_{\mu\nu}=\partial_\mu 
A_\nu - \partial_\nu A_\mu - ig[A_\mu ,A_\nu].$
Also we use the Euclidean convention, for metric 
$g_{\mu\nu}=\textrm{diag}[+,+,+,+]$,
such that our magnetic and electric fields in Minkowski spacetime read 
$G_{ij}=\epsilon_{ijk}B_k$ and $G_{4i}=-iE_i$ respectively. 
In the worldline path integral
formalism the non-Abelian effective action may be written 
as~\cite{PhysRev.80.440,*PhysRev.84.108,
*Schubert200173,*GSCHMIDT1993438,*SCHMIDT199469}
\begin{equation}
\Gamma[A]=\int_{0}^{\infty}\frac{dT}{T}e^{-m^2 T}
\oint\mathcal{D}x\,e^{-\int_{0}^{T}d\tau
\frac{1}{4}\dot{x}^2}\,\mathcal{W}\,.
\label{eq:eff_action_pi}
\end{equation}
Here the coordinate boundary conditions, $x(0)=x(T)=x'$, 
are periodic with path
integral measure $\oint\mathcal{D}x\coloneqq
\int dx'\int\mathcal{D}x$.
$\dot{x}\coloneqq dx/d\tau$. Spacetime indices to be summed over are 
understood and suppressed for readability. The 
path ordered Wilson loop reads
\begin{equation}
\mathcal{W}\coloneqq \tr \mathcal{P} e^{ig\int_{0}^{T}d\tau A\dot{x}}\,.
\label{eq:wilson_loop_def}
\end{equation}

The challenge here in contrast to an Abelian gauge for the application
of steepest descents is a matrix weighted worldline action. However,
with application of a coherent state approach, we will show, 
the Wilson loop may be cast as a path integral with two merits: 
1. The worldline action becomes a c-number amenable to a worldline 
instanton approach. 2. The path-ordering is negated. 
Furthermore, for the most general configuration 
we extend the application of the coherent state formalism to the Wilson 
loop in SL$(2,\mathbb{C})$.

\subsection{Coherent State Formalism in 
SL\boldmath{$(2,\mathbb{C})$}}

To cast the Wilson loop as a coherent state path integral we follow
the approach used in~\cite{PhysRevD.58.105016,Zhang:1990fy},
whereby we extend applicability of the coherent state Wilson loop to
the non-compact group, $G=\,$SL$(2,\mathbb{C})$. The essence
of the approach entails one find a Haar measure leading to a 
resolution of the identity, to which one may insert into the
infinitesimally segmented Wilson loop.

Let us first go over relevant or basic details of $G$; 
$G$ is described through the algebra, 
$[l^i , l^i] = i\epsilon^{ijk}l^k$, $[l^i , k^i] 
= i\epsilon^{ijk}k^k$, and $[k^i , k^i] = -i\epsilon^{ijk}k^k$,
for $l^i = \sigma^i /2$ and $k^i = il^i$ corresponding to the generators of 
SU$(2)$ and SU$(1,1)$ respectively, and so both $l$ and $k$ transform 
as vectors under SU$(2)$. 

One can make use of the coherent state formalism~\cite{Zhang:1990fy} 
through an exploitation of the \textit{one-to-one} correspondence 
between coherent states $| \alpha \rangle$, representing elements of the 
group $G$, and points $\alpha$ in the complex plane.
The map is a continuous manifestation of the topological equivalence 
between these two spaces. In this way, distances are 
determined by the intrinsic metric associated to the inner product, of 
which this Hilbert space is endowed with.

The Lie algebra of $G$, $\mathfrak{sl}(2,\mathbb{C})$, is semi-simple, 
and so it is more convenient to rewrite it in its standard Cartan basis: 
$\{H_\beta , E_\beta, E_{-\beta}\}$, with the usual
diagonal, $H_\beta$, and off-diagonal, $E_\beta$, shift operators.
For the construction of coherent states of a dynamical group $G$, we will 
require a normalized reference state 
$| \Lambda \rangle \in \mathcal{H}^{\Lambda}$ with Hilbert space
$\mathcal{H}^{\Lambda}$ corresponding to the unitary irreducible 
representation of $G$. And we will also require the maximum stability 
subgroup $H$ and coset $G/H$~\cite{Zhang:1990fy}. Here we have
$H=$ SU$(2)$ pointing towards $G/H=$ SL$(2,\mathbb{C})/$SU$(2)$ 
as the target space. The $\mathfrak{sl}(2,\mathbb{C})$ algebra is 
obtained through complexification of SU$(2)$ group algebra, e.g., 
$\mathfrak{su}(2)\otimes i \mathfrak{su}(2)$ and so its
topology is $S^3 \times H^3$.
$G/H$ is topologically equivalent to the upper sheet of
a 3-dimensional mass hyperboloid $H_{3}^{+}$, and is thus endowed with 
a hyperbolic metric as we will see.
The coset element can be generated through the action of the 
displacement operator, with $u\in$ SL$(2,\mathbb{C})$ where we write 
$| \Lambda, u \rangle\coloneqq u| \Lambda \rangle$,
\begin{equation}
| \Lambda, u \rangle = \exp  (\eta_\beta E_\beta -
\bar{\eta}_\beta E_{-\beta})  | \Lambda \rangle\,,
\label{eq:coherent}
\end{equation}
where a sum over the roots of the algebra--above denoted with 
$\beta$--is implicit, with angle $\eta_\beta \in\mathbb{C}$.
The states are in one-to-one correspondence and topologically 
equivalent to $G/H$. Then making use of 
Baker-Campbell-Hausdorff formula one can show the 
displacement operator becomes~\cite{Zhang:1990fy}
\begin{align}
| \Lambda, u \rangle & = e^{ z_\alpha E_\alpha} 
e^{\gamma_\sigma H_\sigma } e^{- \bar{z}_\beta E_{-\beta} }
| \Lambda, z \rangle  \\
& = N(z,\bar{z})^{-1/2} \exp \big(z_\alpha E_{\alpha} \big) 
| \Lambda, z \rangle\,.
\label{eq:BCH}
\end{align}
All powers of $e^{- \bar{z}_\beta E_{-\beta} }$
vanish after acting on the extremal state.
 The action of $e^{\gamma_{\sigma} H_{\sigma}}$ gives rise to the
the normalization factor $N(z,\bar{z})$ identified
as the K{\"a}hler potential $K\coloneqq\log N(z,\bar{z})$.
The invariant metric then can be found as the second derivative of the 
potential $g_{\alpha \beta} = \partial \bar{\partial} K $,
with K{\"a}hler two-form 
$\omega = i g_{\alpha \beta} dz^{\alpha} \wedge d\bar{z}^{\beta}$. And 
the invariant measure is used in the construction of the Haar measure; to 
find the measure it is convenient to make use of a complex projective 
map.

Let us then consider the Cartan decomposition
$\mathfrak{g} = \mathfrak{q} \oplus \mathfrak{p}$ of the 
$\mathfrak{sl}(2,\mathbb{C})$ Lie
algebra, with $\mathfrak{q}$ the Lie algebra of $\mathfrak{su}(2)$ and 
$\mathfrak{p}=\eta_\beta E_\beta - \bar{\eta} E_{-\beta}$
its orthogonal complement, such that
$[\mathfrak{q},\mathfrak{q}] \subset \mathfrak{q}$,
$[\mathfrak{q},\mathfrak{p}] \subset \mathfrak{p}$,
and $[\mathfrak{p},\mathfrak{p}] \subset \mathfrak{q}$
as outlined in~\cite{Zhang:1990fy}.
Then we can see that for non-compact groups, such as our case, the matrix 
representations of generators $R(\mathfrak{q}), R(\mathfrak{p})$ are 
skew-symmetric and symmetric respectively, 
therefore the coset SL$(2,\mathbb{C})$/SU$(2)$ 
is a symmetric space. Let us look at a
matrix representation of the non-compact coset 
group~\cite{hua1963harmonic,*helgason1979differential},
\begin{equation}
\begin{pmatrix}
\sqrt{1+ w\bar{w}} & w \\ \bar{w} & \sqrt{1+ w\bar{w}}\end{pmatrix}
\end{equation}
with $w=\eta \sinh(|\eta|)/|\eta|$.
Likewise, our coset group defined as
the set of all Hermitian two-by-two matrices with determinant one, can 
be parametrized by the convenient global coordinates of $H^{+}_{3}$ as
\begin{equation}
\begin{pmatrix}1 & z \\ 0 & 1\end{pmatrix}
\begin{pmatrix}e^{\gamma/2} & 0 \\ 0 & e^{-\gamma/2}\end{pmatrix}
\begin{pmatrix}1 & 0 \\ \bar{z} & 1\end{pmatrix} =
\begin{pmatrix}e^{\frac{\gamma}{2}} + |z|^2 
e^{-\frac{\gamma}{2}}& z e^{-\frac{\gamma}{2}} \\ 
\bar{z} e^{-\frac{\gamma}{2}} & e^{-\frac{\gamma}{2}}\end{pmatrix}\,.
\label{eq:parametrization}
\end{equation}
Equivalently, one may arrive at this expression by explicitly working from 
the displacement operator, Eq.~\eqref{eq:coherent}.
Equating these two parametrizations for the coset group
leads us to the complex projective map: $z = w(1+w\bar{w})^{1/2}$ and 
$e^{\gamma/2} = (1+w\bar{w})^{-1/2}$.
It follows that any group element $u\in$ SL$(2,\mathbb{C})$ 
(parametrized by coefficients  $\alpha,\beta,\gamma,\delta$) acting on 
a coset element $z$ is a holomorphic M{\"o}bius
transformation, i.e., $T_{u}(z) = (\alpha z + \beta)/(\gamma z + \delta)$. 

Equipped with this chart, we can rewrite the left hand side of 
Eq.~\eqref{eq:parametrization} 
purely in terms of $z$ and express it in the form of Eq.~\eqref{eq:BCH}.
The normalization of these states follows from an explicit computation of 
the exponential operator on the reference state.

\begin{equation}
| \Lambda, u \rangle = \frac{1}{(1 - z\bar{z})^{\Lambda}} 
\exp (z_\alpha E_\alpha) | \Lambda, z \rangle. 
\end{equation}
At this point the K{\"a}hler
potential, and hence its invariant Haar coset metric, 
can be recognized:
\begin{align}
g_{\alpha\bar{\beta}} & = -\Lambda \, \partial \bar{\partial} 
\log(1-z\bar{z}) \\
& = \frac{\Lambda}{(1-z\bar{z})^2}(\delta_{\alpha\bar{\beta}}(1-z\bar{z})
+z^{\alpha}\bar{z}^{\beta})
\end{align}
The Haar measure eventually becomes,
\begin{equation}
d\mu_{\Lambda}(z,\bar{z})
= \frac{2\Lambda+1}{4\pi}\frac{dz d\bar{z}}{(1 - z\bar{z})^{2}}\,,
\label{eq:Haar}
\end{equation}
for the coset SL$(2,\mathbb{C})$/SU($2$), that is the complex 
projection of upper sheet of the 3-dimensional hyperboloid $H_{3}^+$, 
the Poincare disk metric.

We can in this way expand any arbitrary state
$| \Psi \rangle \in \mathcal{H}^\Lambda$ Hilbert space into coherent 
space whose coefficients are smooth functions
in $u$ defined over the coset space. However, due to the 
overcompleteness of coherent states, 
the expansion is not unique. And so the expansion coefficients may be 
deformed. This set of functions will on the other hand allow us to 
construct the basis for function space
L$^2($SL$(2,\mathbb{C})$/SU$(2))$~\cite{Gawedzki:1991yu,
*ISHIBASHI2000149,*TESCHNER1999390}:
\begin{equation}
\mathcal{H} \equiv \mathrm{L}^2(H_{3}^{+})= 
\int^{\oplus}_{s \in \mathbb{R}^{+}} ds s^2 \, 
\mathcal{H}_{-1/2 + is}\,,
\end{equation}
through the principal series decomposition of the group, where the spin 
$j$ takes on complex values, $j=-\frac{1}{2} +is$, 
$s\in\mathbb{R}$, coming from the Casimir having a continuous 
spectrum. For a detailed account on SL$(2,\mathbb{C})$ we refer the 
reader to~\cite{doi:10.1063/1.3533393}.

Equipped with the Haar measure and hence the resolution
of identity we may transform the Wilson loop into a path
integral over coherent states. To cast Eq.~\eqref{eq:wilson_loop_def} into 
its path integral form we must both expand the path
ordered exponential of parallel transporter 
$\mathcal{W}_{L}(\tau_0 ,\tau)$ and evaluate its trace to
recover the Wilson Loop. This is a standard procedure;
we first partition the path $L:\tau_0 \rightarrow \tau$ into infinitesimal 
segments,
\begin{equation}
\mathcal{P} \exp \left[ig \int_{\tau_0}^{\tau}d\tau A\dot{x} \right]= 
\mathcal{P} \prod_{k=0}^{N-1} (1+ig\epsilon A\dot{x})\,,
\end{equation}
where $\epsilon = (\tau-\tau_0)/N$ and take proper limits at the end. To 
evaluate trace of this object we are required to choose a set of states 
resolving the identity 
$1= \int |\Lambda, u_k \rangle d\mu(u_k) \langle \Lambda, u_k|$,
with $u_k$ a coset element at $\tau_k$ and Haar measure given in 
Eq.~\eqref{eq:Haar}, to which we may insert at each partition point. The 
Wilson loop becomes
\begin{equation}
\mathcal{W}=\int\mathcal{D}\mu_{C}\exp\Bigl\{\frac{ig}{2}\oint_{C}
(m^a A^adx + \omega(x)) \Bigr\}\,.
\end{equation}
with $\mathcal{D}\mu$ corresponding to the product of the invariant 
Haar measure per coset element and
$m^a = \langle \Lambda | u(x)T^a u^{-1}(x)| \Lambda \rangle$ and 
$\omega(x) = \langle \Lambda | u(x)d u^{-1}(x)| \Lambda \rangle$.
Let us notice that the argument of the exponential has been brought 
down to Abelian quantities allowing us to apply Stokes' theorem (hence 
coined the NAST as demonstrated for SU$(N)$ 
in~\cite{PhysRevD.58.105016,DIAKONOV1989131,
*10.1143/PTP.104.1189,*KONDO20151}) and express W over
the surface bounded by C, 
\begin{equation}
\mathcal{W}=\oint\mathcal{D}\mu\exp\Bigl\{\frac{ig}{2}\int_{0}^{T}d\tau\,
\tr [\sigma_{3}(uA\dot{x}u^{-1}+\frac{i}{g}u\dot{u}^{-1})]\Bigr\}\,.
\label{eq:wilson_loop_coherent}
\end{equation}
Having demonstrated the Wilson loop may be represented in 
SL$(2,\mathbb{C})$
as a path integral over a complex isospin, let us now show how periodic
solutions to Wong's equations in Euclidean spacetime are non-Abelian
worldline instantons. In evaluating the trace we adopt coherent states, 
which are equipped with properties that will prove useful to explore more 
exotic field configurations. Coherent states, in terms of the language of 
group theory, are embedded in a topologically nontrivial geometrical 
space facilitating the perfect machinery to probe instanton and
other nontrivial configurations.

\subsection{Wong's Equations}

Wong's equations follow as classical equations of motion of the
worldline action. Let us however first arrange terms in the action 
to more suitably elicit their connection to worldline instantons; to
accomplish this we first take the Schwinger proper time integral, $T$,
as was done in~\cite{AFFLECK1982509,PhysRevD.72.105004}.
First, using the coherent state represented Wilson loop in 
Eq.~\eqref{eq:wilson_loop_coherent}, we have for the effective action, 
Eq.~\eqref{eq:eff_action_pi}, after the substitution, 
$\tau\rightarrow T\tau'$
\begin{align}
&\Gamma[A]=\int_{0}^{\infty}\frac{dT}{T}
\oint\mathcal{D}x\mathcal{D}\mu\,e^{-S}\\
&S=\int_{0}^{1}d\tau\Bigl\{m^2T+\frac{\dot{x}^{2}}{4T}
-\frac{ig}{2}\tr\bigl[\sigma_{3}\bigl(uA\dot{x}u^{-1}
-\frac{i}{g}\dot{u}u^{-1}\bigr)\bigr]\Bigr\}\,.
\label{eq:wl_action}
\end{align}
Let us evaluate the proper time integral through steepest 
descents--or rather the Laplace method.
We expand $T=T_{n}+K\exp(i\alpha_{n})$ for stationary points,
$T_{n}$, about $K\in[0,\infty)$, with phase 
$\alpha_{n}=\pi/2-1/2\arg(f''(T_{n}))$,
for $f(T)=-m^{2}T-\frac{1}{4T}\int_{0}^{1}d\tau\dot{x}^{2}$. 
We also have that $f''(T)=-\frac{1}{2T^{3}}\int_{0}^{1}d\tau\dot{x}^{2}$.
However, in what follows we will confine our attention to the case of
only real and positive $T_n$, and hence the phase factor 
$\exp(i\alpha_n) =1$. We find for the stationary points
\begin{equation}
T_{n}^{2}=\frac{1}{4m^{2}}\int_{0}^{1}d\tau\,\dot{x}^{2}\,,
\label{eq:stationary_point}
\end{equation}
whose $n$ minima can be had alongside steepest descents in coordinate
and isospin space. The proper time integral becomes
\begin{equation}
\int_{0}^{\infty}\frac{dT}{T}e^{-m^{2}T
-\frac{1}{4T}\int_{0}^{1}d\tau\dot{x}^{2}}\approx \sum_{n}
\sqrt{\frac{\pi T_{n}}{4m^{2}}}e^{-2m^{2}T_{n}}\,.
\end{equation}
Importantly, that steepest descents be valid a large mass and or weak
fields\textemdash characterized by $\sqrt{\int_{0}^{1}d\tau\dot{x}^{2}}$
\cite{AFFLECK1982509,PhysRevD.72.105004}\textemdash must be 
assumed, i.e., $m^{2}T_{n}\gg1$. 
The criteria is safely met for a low probability
of pair production occurrence. 
Furthermore, while we solve Wong's equations
for complex trajectories, we will find that only real periodic paths
will contribute to pair production for the fields examined. 
Last, the Schwinger effect is
dominated by an exponential suppression, 
therefore we ignore prefactor
terms. 
Let us then turn our attention to the worldline action, 
Eq.~\eqref{eq:wl_action}, for a given stationary 
point in proper time, Eq.~\eqref{eq:stationary_point}, 
\begin{equation}
S_{n}=m\sqrt{\int_{0}^{1}d\tau\dot{x}^{2}}
-\frac{ig}{2}\int_{0}^{1}d\tau\,
\tr\bigl[\sigma_{3}\bigl(uA\dot{x}u^{-1}-
\frac{i}{g}\dot{u}u^{-1}\bigr)\bigr]\,.
\label{eq:worldline_action}
\end{equation}
Classical solutions of the above action lead to Wong's equations, and 
with the Euclidean periodicity criteria, also lead to worldline instantons. 
Let us begin with the gauge element, $u$, leading to Wong's equation 
describing the precession of isospin.

One may straightforwardly apply the Euler-Lagrange equations to the
action in the gauge element through an introduction of an infinitesimal
angle to the gauge element as outlined 
in~\cite{PhysRevD.17.3247}. Let us assume
the gauge element can be parameterized by the independent set of 
variables given by $\Theta$; then we have for 
$\bar{\sigma}_{i}=(\vec{\sigma},i\vec{\sigma})$,
\begin{align}
u(\Theta(\beta))&\coloneqq e^{\frac{i}{2}\beta_{i}\bar{\sigma}_{i}}
u(\Theta)\,,\\
\frac{\partial}{\partial\beta_{i}}u(\Theta(\beta))
\Big|_{\beta=0}&=\frac{\partial\Theta_{j}}{\partial\beta_{i}}
\frac{\partial}{\partial\Theta_{j}}u(\Theta(\beta))\Big|_{\beta=0}\,,
\end{align}
where the define $M_{ij}\coloneqq
\frac{\partial\Theta_{j}}{\partial\beta_{i}}|_{\beta=0}$.
And we find
\begin{equation}
\frac{\partial}{\partial\Theta_{j}}u(\Theta)=M_{ij}^{-1}
\frac{i}{2}\bar{\sigma}_{i}u(\Theta)\,.
\end{equation}
$M$ must have a non-zero determinant owing to the linear 
independence of $\bar{\sigma}$. 
Then, using the fact that $\dot{u}^{-1}=-u^{-1}\dot{u}u^{-1}$
and likewise $\frac{\partial}{\partial\Theta_{i}}u^{-1}=-u^{-1}
(\frac{\partial}{\partial\Theta_{i}}u)u^{-1}$,
we find for parameter, $\Theta_i$,
\begin{equation}
\frac{\delta S}{\delta\Theta_{i}} 
=\frac{i}{2}M_{ji}^{-1}\textrm{tr}
\bigl[u^{-1}\bar{\sigma}_{j}u
\bigl\{[I,igA\dot{x}]+[I,u^{-1}\dot{u}]\bigr\}\bigr]\,,
\end{equation}
Here we have introduced the isospin variable,
\begin{equation}
I=\frac{1}{2}u^{-1}\sigma_{3}u\,.
\end{equation}
Then using that
$\dot{I} =[I,u^{-1}\dot{u}]$,
we can find for the isospin portion of Wong's equations as
\begin{equation}
\dot{I}=[igA\dot{x},I]\,.
\label{eq:isospin_em}
\end{equation}

Let us digress on some properties of the isospin. It is conserved
as $\tr I^{2}=(1/2)I_{i}I_{i}=1/2$, where 
$I=(I_{i}/2)\sigma_{i}$
and $I_{i}=\tr [\sigma_{i}I]$.
We can see therefore that the isospin performs a 
precessional motion~\cite{Wong:1970fu} despite its complex 
nature. 

One may also write the isospin equation, Eq.~\eqref{eq:isospin_em}, as 
$[(1/2)\sigma_{3},uigA\dot{x}u^{-1}+\dot{u}u^{-1}]=0$,
and hence we can see that a solution to the isospin equation of motion
is analogous to finding a gauge transformation of the Hamiltonian
$igA\dot{x}$ such that the transformed Hamiltonian becomes diagonal.
Alternatively, a solution to the isospin equation of motion furnishes
such a gauge. An exact solution can be had in principal such that
the off diagonal pieces of the geometric phase $\dot{u}u^{-1}$ cancel
with those of the $uigA\dot{x}u^{-1}$ term. We, however, exploit
an adiabatic approximation giving way to the familiar 
Berry's phase~\cite{doi:10.1098/rspa.1984.0023} to arrive
at a solution of Eq.~\eqref{eq:isospin_em}.

Wong's other equation is the non-Abelian equivalent of the Lorentz
force equation. 
This can be found straightforwardly by minimizing
the worldline action, Eq.~\eqref{eq:worldline_action}, with respect to
the coordinate $x$, making use of the isospin equation, 
Eq.~\eqref{eq:isospin_em},
and the fact that $\dot{x}^{2}$ is a constant owing to the anti-symmetric 
field strength tensor. We find for the 
Lorentz force portion of Wong's equations as
\begin{equation}
\ddot{x}_\mu=-\frac{ig|\dot{x}|}{m}\tr [IG_{\mu\nu}]\dot{x}_{\nu}\,.
\label{eq:lorentz_em}
\end{equation}
$|\dot{x}|\coloneqq \sqrt{\dot{x}^2}$. Eqs.~\eqref{eq:isospin_em} 
and~\eqref{eq:lorentz_em} make up Wong's
equations. Their solutions in Euclidean spacetime about periodic 
boundary conditions are worldline instantons predicting a particle 
antiparticle tunneling from the QFT vacuum. Here, we focus on the 
dominant exponential suppression and confine our attention to the 
classical worldline action, whose form is
\begin{equation}
S_{n}=m|\dot{x}|-\int_{0}^{1}d\tau\,\textrm{tr}\bigl[IigA\dot{x}
+\frac{1}{2}\sigma_3\dot{u}u^{-1}\bigr]\,,
\label{eq:S_n}
\end{equation}
for instanton number, $n$, governed by Eqs.~\eqref{eq:isospin_em} 
and~\eqref{eq:lorentz_em}. Let us next look at the simplest case of
non-Abelian fields that posses non-zero Chern-Pontryagin density; 
these are homogeneous parallel fields. This step is instructive in 
that we can check the validity of the extension of the worldline instanton 
method to non-Abelian fields. But moreover, a key difference 
between the electric and magnetic field applicability to the WI method is 
revealed, which is beneficial to expose in a comparatively simple setting.
The non-Abelian WI method devised above is new,
and let us remark that one may also arrive at our results starting from an 
auxilary field approach~\cite{BARDUCCI1981141,*Bastianelli_2013,
*Bastianelli:2015iba,Corradini:2016czo,
*Edwards:2016acz,*Ahmadiniaz:2018olx} to a complexification of the 
group to SU$(2)^{\mathbb{C}}$.

\subsection{Homogeneous Abelian-like Parallel Fields}
\label{sec:homo}

Let us examine homogeneous parallel fields in the $\hat{x}_3$ direction,
which are in an Abelian-like representation:
\begin{align}
G_{12} &=B\sigma_{3}\,,\quad G_{34}=iE\sigma_{3}\,, 
\label{eq:homo_G}\\
A_{2} &=Bx_{1}\sigma_{3}\,,\quad A_{4}=iEx_{3}\sigma_{3}\,.
\label{eq:homo_fields}
\end{align}
These fields correspond to SU$(2)$ parallel fields, $\vec{B}=B\hat{x}_3$
and $\vec{E}=E\hat{x}_3$, in Minkowski spacetime; 
here they are in SL$(2,\mathbb{C})$. For Abelian-like,
(here proportional to $\sigma_{3}$), non-Abelian fields the isospin, 
according to Eq.~\eqref{eq:isospin_em},
takes a trivial solution. 
Note that for fields not proportional to $\sigma_{3}$, the
Schwinger effect characteristics differ markedly~\cite{BROWN1979285}.
Since $[\sigma_{3},A_{\mu}]=0$ we may select
a propertime independent gauge; 
let us take $u_{-,+}=i\sigma_{1,3}$
and $u_{-,+}^{-1}=-i\sigma_{1,3}$. 
Then for $I^H_{\pm}=\frac{1}{2}u_{\pm}^{-1}\sigma_{3}u_{\pm}$,
we find two independent solutions for the isospin as
\begin{equation}
I^H_\pm=\pm\frac{1}{2}\sigma_{3}\,.
\label{eq:homo_isospin}
\end{equation}
Variables of solutions in a homogeneous parallel field background
are affixed with superscript, $H$, to contrast later solutions in a
BPST instanton background, $I$, and a complex equivalent, $CI$.

The Lorentz force equation, Eq.~\eqref{eq:lorentz_em} reduces to
\begin{align}
\ddot{x}^H_{\pm\,1}&=\mp\frac{ig|\dot{x}|}{m}B\dot{x}^H_{\pm\,2}\,,
\quad\ddot{x}^H_{\pm\,2}=\pm\frac{ig|\dot{x}|}{m}B\dot{x}^H_{\pm\,1}\,,\\
\quad\ddot{x}^H_{\pm\,3}&=\pm\frac{g|\dot{x}|}{m}E\dot{x}^H_{\pm\,4}\,,
\quad\ddot{x}^H_{\pm\,4}=\mp\frac{g|\dot{x}|}{m}E\dot{x}^H_{\pm\,3}\,.
\end{align}
As we are looking for only solutions that yield real and positive
$|\dot{x}|$ we examine only solutions corresponding to the electric
field. One can readily see that the coordinates associated with the 
magnetic field will be periodic about an imaginary argument for a 
hyperbolic sinusoidal function; whereas the electric field periodicity
is governed by a real argument in a sinusoidal function.
No simultaneous solutions for
both the electric and magnetic parts exist. Physically, this stems from
the fact that a sole magnetic field cannot elicit Schwinger pair 
production. Therefore one may take for the magnetic field
coordinates, $x^H_1$ and $x^H_2$, a trivial constant value.
Whereas for the electric field coordinates,
we find for the non-Abelian WIs the same as in the Abelian 
case \cite{AFFLECK1982509,PhysRevD.72.105004}:
\begin{align}
x^H_{+\,3}(\tau)&=x^H_{-\,4}(\tau)
=\frac{m}{gE}\sin\bigl(\frac{g|\dot{x}|E}{m}\tau\bigr)\,,\nonumber \\
x^H_{+\,4}(\tau)&=x^H_{-\,3}(\tau)
=\frac{m}{gE}\cos\bigl(\frac{g|\dot{x}|E}{m}\tau\bigr)\,,
\end{align}
with two distinct WIs circling either clockwise or 
counterclockwise according to the isospin, 
Eq.~\eqref{eq:homo_isospin}. 
And to satisfy
periodic boundary conditions we must have similarily
\begin{equation}
|\dot{x}|=\frac{2n\pi m}{gE}\quad\forall n\in\mathbb{Z}^{+}\,.
\end{equation}
The WIs indicate an exponential suppression of
Schwinger pair production according to Eq.~\eqref{eq:S_n} of
\begin{equation}
S_{n\,\pm}=\frac{\pi nm^{2}}{gE}\,,
\label{eq:sn_homo}
\end{equation}
which is the same for either $I_{\pm}$. The exponential suppression 
follows the exact solution, Eq.~\eqref{eq:homo_exact}, as is calculated in
the Appendix.
Having illustrated the simplest case of non-Abelian pair production 
via the worldline instanton method, let us address the case of pair 
production in a YMI background; we will find as expected 
pair production is absent.

\section{Pair Production (or lack thereof) in a 
BPST Instanton Background}
\label{sec:pp_bpst}

To provide a convenient point of comparison for later discussions,
let us write out the BPST instanton (YMI) field, 
$A^{I}$, in the regular gauge,
\begin{equation}
A_{\mu}^I (x)=\frac{i}{g}\frac{x^{2}}{x^{2}+R^{2}}
G(\hat{x})^{\dagger}\partial_{\mu}G(\hat{x})\,,
\label{eq:instanton}\
\end{equation}
where $R$ denotes the size of the instanton, but has no effect on 
the topological winding number. The anti-YMI,
$A^{\bar{I}}$, can be found with the replacement 
$G\rightleftharpoons G^\dagger$. We use the conventions
of~\cite{Bohm:2001yx,*srednicki_2007}.
Then for the
gauge element $G(\hat{x})=\hat{x}_{\mu}\bar{\sigma}_{\mu}$, where
$\sigma_{\mu}\coloneqq(i\vec{\sigma},1)$ and
$\bar{\sigma}_{\mu}\coloneqq(-i\vec{\sigma},1)$,
the gauges read
\begin{equation}
A_\mu^I (x)=\frac{1}{g}\frac{2}{x^2+R^2}\sigma_{\mu\nu}x_\nu\,,\;
A_{\mu}^{\bar{I}}(x)=\frac{1}{g}\frac{2}{x^{2}+R^{2}}
\bar{\sigma}_{\mu\nu}x_{\nu}\,,
\end{equation}
where 
$\sigma_{\mu\nu}\coloneqq\frac{1}{4i}
[\sigma_{\mu}\bar{\sigma}_{\nu}-\sigma_{\nu}\bar{\sigma}_{\mu}]$ and
$\bar{\sigma}_{\mu\nu}:=\frac{1}{4i}
[\bar{\sigma}_{\mu}\sigma_{\nu}-\bar{\sigma}_{\nu}\sigma_{\mu}]$.
And the field strength tensors are of course
\begin{equation}
G_{\mu\nu}^{I}=
-\frac{4R^{2}}{g(x^{2}+R^{2})^{2}}\sigma_{\mu\nu}\,,\;
G_{\mu\nu}^{\bar{I}}=
-\frac{4R^{2}}{g(x^{2}+R^{2})^{2}}\bar{\sigma}_{\mu\nu}\,.
\end{equation}

For the calculations that follow it is convenient  
to use a matrix form for Lorentz indices; contractions are understood and 
$x^T=(x_1,x_2,x_3,x_4)$.
Let us show this for the `t Hooft symbols. For $\sigma_{\mu\nu}
=\frac{1}{2}\eta_{a\mu\nu}\sigma_{a}$ and 
$\bar{\sigma}_{\mu\nu}=\frac{1}{2}\bar{\eta}_{a\mu\nu}\sigma_{a}$, we
have for the symbols,
$\eta_{a\mu\nu}=\varepsilon_{a\mu\nu4}
+\delta_{a\mu}\delta_{\nu4}-\delta_{a\nu}\delta_{4\mu}$ and
$\bar{\eta}_{a\mu\nu}=\varepsilon_{a\mu\nu4}
-\delta_{a\mu}\delta_{\nu4}+\delta_{a\nu}\delta_{4\mu}$.
The `t Hooft symbols can be shown to satisty several relationships. They
are antisymmetric: $\eta_{i}^{T}=-\eta_{i}$ and 
$\bar{\eta}_{i}^{T}=-\bar{\eta}_{i}$. And also, since the symbols transform 
under SO$(4)=$ SU$(2) \,\otimes\,$SU$(2)$, we can find the following:
\begin{equation}
\eta_{i}\eta_{j}=-[\delta_{ij}+\varepsilon_{ijk}\eta_{k}]\,,\quad
\bar{\eta}_{i}\bar{\eta}_{j}=-[\delta_{ij}+\varepsilon_{ijk}\bar{\eta}_{k}]\,.
\end{equation}

Let us go ahead and express Wong's equations in the above form 
for the YMI, Eq.~\eqref{eq:instanton};
they are
\begin{align}
\dot{I}^I_c&=-\frac{2}{x^{I\,2}+R^{2}}
\varepsilon_{abc}\dot{x}^{I\,T}\eta_{a}x^I I^I_{b}\,,
\label{eq:isospin_matrix}\\
\ddot{x}^I&=\frac{i|\dot{x}^I|}{m}
\frac{2R^{2}}{(x^{I\,2}+R^{2})^{2}}\eta\cdot I^I\dot{x}^I\,.
\end{align}
Before explicit computation, we can see that as was the case for the
magnetic fields in the previous section, see 
Sec.~\ref{sec:homo} and Eq.~\eqref{eq:homo_G},
the Lorentz force equation has a real field strength argument, in contrast
to the electric field. Moreover, eigenvalues of 
$\textrm{tr}[I^IG^I_{\mu\nu}]$
are all real, and hence only project magnetic parts. We can explore this
more deeply with the aid of a large instanton limit and through fixing the 
isospin in the direction of the gauge.
Let us also point out that Wong's equation in the YMI have been 
studied in~\cite{10.1143/PTP.62.544,*montgomery}, however our 
WI approach as well as calculation technique are new.

\subsection{Adiabatic Theorem and the Large Instanton}
\label{sec:aa_YMI}

Let us first evaluate the isospin equation of motion.
Consider for fictitious Hamiltonian, $H=igA\dot{x}$, 
the isospin equation
of motion provided by 
$[\frac{1}{2}\sigma_{3},uHu^{-1}+\dot{u}u^{-1}]=0$; see also
Eq.~\eqref{eq:isospin_em}. One can immediately see a solution is 
provided by the selection of
a gauge element, $u$, such that $H$ takes a diagonal form, and that 
off-diagonal parts of the 
geometric phase, $\dot{u}u^{-1}$, may be ignored. 
This is an adiabatic theorem leading
to Berry's phase(s)~\cite{doi:10.1098/rspa.1984.0023}, 
one for each monopole singularity governing 
level crossing in $H$. We will demonstrate shortly that the adiabatic
approximation is equivalent to (complex) circular solutions for
the WIs, as we calculated in Sec.~\ref{sec:homo}, and the importance
of which we highlighted in Sec.~\ref{sec:mink_pp}. 
Moreover, there, for the homogeneous fields circular solutions to 
Wong's equations were found, and homogeneous fields were also
found to be the limiting form in the large instanton limit, 
Eq.~\eqref{eq:large_instanton}. 
Thus the adiabatic theorem and large instanton limits go hand in hand. 
We take for the gauge element, $u^I$, such that
\begin{equation}
u^IA^{I}\dot{x}^I(u^I)^{-1}
=\sqrt{A^{I\,a}\dot{x}^IA^{I\,a}\dot{x}^I}\sigma_3\,,
\label{eq:aa_gauge}
\end{equation}
and likewise for $A^{\bar{I}}$. Let us just treat the YMI from this point, 
and report on the anti-YMI below. One can see that the above 
corresponds to
\begin{equation}
I^{I}_a=\frac{\dot{x}^{I\,T}\eta_{a}x^I}
{\sqrt{\dot{x}^{I\,T}\eta_{a}x^I\dot{x}^{I\,T}\eta_{a}x^I}}\,.
\label{eq:isospin_ap}
\end{equation}
We can see the isospin is fixed in the direction of the gauge field,
whose magnitude is always unity since the isospin is an element 
of the coset SU$(2)/$U$(1)$ and is tracing out a point on the surface 
of the unit sphere.
One can see an adiabatic approximation entails isospin also to be independent of proper time, since according to 
Eq.~\eqref{eq:isospin_matrix}, $\dot{I}^I=0$. Then consider
Eq.~\eqref{eq:isospin_ap}, since $\eta_a$ are all antisymmetric
tensors we can see that to satisfy $\dot{I}^I=0$ we must have that
\begin{equation}
x^I\propto \ddot{x}^I\,.
\label{eq:adiabatic_constrait}
\end{equation}

We also take the large instanton limit $R^2\gg x^2$; then one 
can see the field strength tensor takes on the following form
(c.f., Sec~\ref{sec:mink_pp})
\begin{equation}
\ddot{x}^I\approx \frac{2i|\dot{x}^I|}{mR^2}
\eta\cdot I^I\dot{x}^I\,.
\label{eq:lorentz_large}
\end{equation}
It proves convenient to evaluate the above using projection 
operators of the field strength tensor, as one might do 
for the Abelian Lorentz force equation~\cite{Fradkin_1978}.
However, due to the self-duality of the YMI, we find
a compact form for the operators, only linear in $\tr [I^IG^I]$; they are
\begin{equation}
P_{-}^{I} \coloneqq\frac{1}{2}(1+i\eta\cdot I^I)\,,\quad 
P_{+}^{I}\coloneqq\frac{1}{2}(1-i\eta\cdot I^I)\,.
\end{equation}
Apart from a Lorentz index assignment there is no real identification
for electric and magnetic projection operators; they both project out
magnetic field components. Some useful properties of the 
projection operators include idempotency, a completeness, and an
orthogonality between unlike projectors:
\begin{equation}
P^{I\,2}_{\pm}=P^{I}_{\pm}\,,\quad P_{-}^{I}+P_{+}^{I}=1\,,\quad
P_{\pm}^{I}P_{\mp}^{I}=0\,.
\end{equation}
Last, the projection operators project their respective 
eigenvalues of the field strength tensor,
\begin{equation}
\eta\cdot IP_{-}^{I}=-iP_{-}^{I}\,,\quad \eta\cdot IP_{+}^{I}=iP_{-}^{I}\,.
\label{eq:projection_eigen}
\end{equation}
One can use the projection operators to decouple the isospin
from the Lorentz force equation.

Let us use the projection operators to separate the coordinates
into two parts
\begin{equation}
x^I_+\coloneqq P_+^I x^I\,,\quad x^I_-\coloneqq P_-^I x^I\,.
\end{equation}
Note also that the projection operators under interchange of the 
Lorentz indicies satisfy the following relationship: 
$P_{I\,\pm}^{T}=P_{I\,\mp}$. Hence, we would have 
$x_\pm^{I\,T} = x^{I\,T}P^I_\mp$. Furthermore, under the
adiabatic approximation we found the isospin becomes propertime
independent. Therefore, using Eq.~\eqref{eq:projection_eigen}, 
we find the Lorentz force equation decouples as
\begin{equation}
\ddot{x}^I_+=-\frac{2|\dot{x}|}{mR^2}\dot{x}^I_+\,,\quad
\ddot{x}^I_-=\frac{2|\dot{x}|}{mR^2}\dot{x}^I_-\,.
\end{equation}
One can readily find the solutions to the above as
\begin{align}
x_{+}^I(\tau)&=\frac{mR^{2}}{2|\dot{x}^I|}\bigl[1-\exp\bigl(
-\frac{2|\dot{x}^I|}{mR^{2}}\tau\bigr)\bigr]\dot{x}^I_{+}(0)+x^I_{+}(0)\,,\\
x^I_{-}(\tau)&=\frac{mR^{2}}{2|\dot{x}^I|}\bigl[\exp\bigl(
\frac{2|\dot{x}^I|}{mR^{2}}\tau\bigr)-1\bigr]\dot{x}^I_{-}(0)+x^I_{-}(0)\,,
\end{align}
that can be combined giving one
\begin{equation}
x^I(\tau)=\frac{mR^{2}}{2|\dot{x}^I|}\Bigl[\sinh\bigl(
\frac{2|\dot{x}^I|}{mR^{2}}\tau\bigr) + \cosh\bigl(
\frac{2|\dot{x}^I|}{mR^{2}}\tau\bigr)i \eta\cdot I^I\Bigr]\dot{x}^I(0)\,,
\label{eq:ymi_soln}
\end{equation}
with $I^I$ given by Eq.~\eqref{eq:isospin_ap}. We have also applied 
the constraint given in Eq.~\eqref{eq:adiabatic_constrait}; this gives
$x^I(0)=i\frac{mR^{2}}{2|\dot{x}^I|}\eta\cdot I^I\dot{x}^I(0)$. 

As anticipated,
one can readily see that to satisfy the periodicity requirement, 
$x^I(0)=x^I(1)$, one must have
\begin{equation}
|\dot{x}^I|=i mR^2\pi n\quad \forall\, n\in \mathcal{Z}^+\,, 
\label{eq:cymi_criteria}
\end{equation}  
in contradiction to the requirement of a real stationary point,
Eq.~\eqref{eq:stationary_point}, and hence real worldline action,
Eq.~\eqref{eq:S_n}. And thus, there can be no pair production.
The YMI acts like a magnetic field, as we saw in Sec.~\ref{sec:homo},
which does not give rise to the Schwinger effect. Also, evaluation of the
anti-YMI would result in Eq.~\eqref{eq:cymi_criteria} as well.

Let us look at the worldline action.
Using Eq.~\eqref{eq:ymi_soln}, one can find that
$\int_{0}^{1}d\tau\,\textrm{tr}\bigl[I^IigA^I\dot{x}^I\bigr]=
(1/2) m|\dot{x}^I|$. Also, the Berry's phase term in 
Eq.~\eqref{eq:S_n}, given by $\dot{u}^I(u^I)^{-1}$ may also only
introduce a trivial factor of $4\pi i$ into the worldline action.
Thus the exponential suppression goes as $(1/2) i m^2R^2\pi n$,
c.f., Eq.~\eqref{eq:sn_homo}.
If one were to have real electric fields, one would expect a similar but
real quantity. Then, let us explore just such a scenario.

\section{Pair Production in a Complex BPST Instanton Background}
\label{sec:pp_cbpst}

Above we demonstrated no pair production could occur in a 
BPST instanton (YMI),
and then (as aluded to before) one may ask for what topologically
non-trivial background fields could one see pair production.
In the homogeneous parallel field case, Sec.~\ref{sec:homo}, a real 
Minkowski electric field was needed for 
the vacuum instability to be present. Furthermore, in 
Sec.~\ref{sec:mink_pp} it was shown in a large instanton limit
the YMI resembled in Minkowski spacetime non-Abelian
homogeneous imaginary electric fields and real magnetic fields. An
intuitive extension of the YMI that might furnish pair 
production then would be to seek field configurations in which both
real electric and magnetic non-Abelian fields are present. We
construct such a background field here.

Let us however point out that Minkowski Yang-Mills solutions with
topology do exist~\cite{LUSCHER1977321,
*PhysRevD.16.3015,PhysRevD.47.5551,PhysRevD.17.486}, 
whose effect on the anomaly 
have been studied~\cite{PhysRevD.51.4561,*KHOZE1995270}.
But are, however, analytically cumbersome for our purposes. 
The importance of parallel real fields in Minkowski space for 
Yang Mills tunneling is stressed in~\cite{PhysRevD.17.486}.
Let us also note that our gauge construction is similar in objective 
as the one demonstrated in~\cite{Peccei}, and indeed we find here too
that the Yang Mills equations of motion are not satisfied.
Finally, we remark that solutions to the Yang Mills equations of motion in Minkowski spacetime need not have integer Chern-Simons number~\cite{PhysRevD.47.5551}.

\subsection{A Complex BPST Instanton}
The desired background field can be had from a simple identification:
Whereas a BPST instanton (YMI) represents a solution with 
winding number
difference in Euclidean time, to wit, $x_{4}\rightarrow-\infty$ to 
$x_{4}\rightarrow\infty$, the desired background field stipulates
a winding number difference in Minkowski real-time for
$x^{0}\rightarrow-\infty$ to $x^{0}\rightarrow\infty$ such that
\begin{align}
\Delta N_{\textrm{Mink}}&=N_{\textrm{Mink}}(x^{0}\rightarrow\infty)
-N_{\textrm{Mink}}(x^{0}\rightarrow-\infty)\nonumber\\
&=\int dx^{0}d^{3}x\,\frac{g^{2}}{16\pi^{2}}
\textrm{tr}[G_{\textrm{Mink}}\widetilde{G}_{\textrm{Mink}}]\,.
\label{eq:real_time_winding}
\end{align}
Let us define the field under a Minkowski metric,
$g_{\mu\nu}=\textrm{diag}(-,+,+,+)$, with map
\begin{align}
\mathcal{G}_{\textrm{Mink}}(x_{0},\vec{x}) 
&=\frac{1}{\rho}(x_{0}-i\vec{x}\cdot\vec{\sigma})\in \text{SU}(2)\,,\\
\mathcal{G}_{\textrm{Mink}}^{-1}(x_{0},\vec{x}) 
&=\frac{1}{\rho}(x_{0}+i\vec{x}\cdot\vec{\sigma})\,,
\end{align}
with $\rho=\sqrt{x_{\mu}x^{\mu}+2x_0^2}$. 
$\mathcal{G}_{\textrm{Mink}}$ has the topology of $S^3$,
however unlike the map in the YMI, Eq.~\eqref{eq:instanton},
$\mathcal{G}_{\textrm{Mink}}$ cannot map from a vacuum $S^3$;
as we will demonstrate our fields are not classical Yang Mills solutions.
Also, $\rho$, is not a Minkowski four-vector of unit length.
Last, one may indeed envision $\mathcal{G}_{\textrm{Mink}}$ as living
in a Minkowski four dimensional cyclinder with end caps in real time
at $N_{\textrm{Mink}}(x^{0}\rightarrow\pm\infty)$, and disappearing 
at spatial infinity.

Using the above one can construct the 
following gauge connection with $\Delta N_\textrm{Mink}=1$,
\begin{equation}
A_{\textrm{Mink}\,\mu}^{CI}(x_0,\vec{x})=
\frac{i}{g}f(x_{0},\vec{x})\mathcal{G}_{\textrm{Mink}}^{-1}
(x_{0},\vec{x})\partial_{\mu}
\mathcal{G}_{\textrm{Mink}}(x_{0},\vec{x})\,.
\label{eq:a_mink}
\end{equation}
Here $f$ is a function such that $f(0)=0$ and 
$f(\rho\rightarrow\infty)=1$,
The above we can see is in precise analogy to the YMI 
with $x_{4}\rightarrow x_{0}$, and thus analogous arguments hold 
here as well. Notably Eq.~\eqref{eq:a_mink} interpolates winding 
numbers at asymptotic real times. The field is also localized in real 
time and space.

WIs have an intuitive periodic structure in Euclidean
time, therefore let us examine the above field in a Euclidean metric.
This provides the additional benefit of contrast with the YMI. 
In a Euclidean metric sense, we refer to the following
solutions as Complex (anti) Yang Mills Instantons (CYMI). For Wick
rotation, $x_{0}=ix_{4}$, we have 
\begin{align}
\mathcal{G}(x)&=\frac{1}{\rho}(ix_{4}-i\vec{x}\cdot\vec{\sigma})
=\hat{x}_{\mu}\bar{\sigma}_{\mu}^{C}\in \text{SL}(2,\mathbb{C})\,,\\
\mathcal{G}^{-1}(x)&=\frac{1}{\rho}(ix_{4}+i\vec{x}\cdot\vec{\sigma})
=\hat{x}_{\mu}\sigma_{\mu}^{C}\,,
\end{align}
with now $\rho=\sqrt{x^{2}-2x_{4}^{2}}$, and also we have
$\sigma_{\mu}^{C}\coloneqq(i\vec{\sigma},i)$ and
$\bar{\sigma}_{\mu}^{C}\coloneqq(-i\vec{\sigma},i)$.
Then for
$A_{\mu}^{CI}(x)=(i/g)f(\rho)
\mathcal{G}(\hat{x})^{-1}\partial_{\mu}\mathcal{G}(\hat{x})$,
one can find
\begin{align}
&G_{\mu\nu}^{CI}(x)=\frac{2}{g\rho^{2}}
\Bigl\{\bigl[\partial_{\mu}f+\frac{2}{\rho}(f^{2}-f)\partial_{\mu}\rho\bigr]
\sigma_{\nu\sigma}^{C}x_{\sigma}\nonumber \\ 
&-\bigl[\partial_{\nu}f+\frac{2}{\rho}(f^{2}-f)\partial_{\nu}\rho\bigr]
\sigma_{\mu\sigma}^{C}x_{\sigma}\Bigr\}
+\frac{4}{g\rho^{2}}(f^{2}-f)\sigma_{\mu\nu}^{C}\,,
\end{align}
where we have defined
\begin{equation}
\sigma_{\mu\nu}^{C}\coloneqq\frac{1}{4i}
[\sigma_{\mu}^{C}\bar{\sigma}_{\nu}^{C}
-\sigma_{\nu}^{C}\bar{\sigma}_{\mu}^{C}]\,,\quad
\bar{\sigma}_{\mu\nu}^{C}\coloneqq\frac{1}{4i}
[\bar{\sigma}_{\mu}^{C}\sigma_{\nu}^{C}
-\bar{\sigma}_{\nu}^{C}\sigma_{\mu}^{C}]\,.
\end{equation}
Then in analogy to the YMI we seek a solutions such that 
$\partial_{\mu}f+\frac{2}{\rho}(f^{2}-f)\partial_{\mu}\rho=0$;
this can be found as $f=\rho^{2}/(\rho^{2}+R^{2})$, and hence
we have for the CYMI and anti-CYMI,
\begin{align}
A_{\mu}^{CI}(x)&=\frac{2}{g}
\frac{\sigma_{\mu\nu}^{C}x_{\nu}}{\rho^{2}+R^{2}}\,,\quad
G_{\mu\nu}^{CI}(x)=-\frac{4}{g}
\frac{R^{2}}{(\rho^{2}+R^{2})^{2}}\sigma_{\mu\nu}^{C}\,,\\
A_{\mu}^{C\bar{I}}(x)&=\frac{2}{g}
\frac{\bar{\sigma}_{\mu\nu}^{C}x_{\nu}}{\rho^{2}+R^{2}}\,,\quad
G_{\mu\nu}^{C\bar{I}}(x)=-\frac{4}{g}
\frac{R^{2}}{(\rho^{2}+R^{2})^{2}}\bar{\sigma}_{\mu\nu}^{C}\,.
\end{align}
$G^{C\bar{I}}$ describes a field configuration in which 
$\Delta N_\textrm{Mink}=-1$.

One can readily demonstrate the sought electric and magnetic
field decomposition of the CYMI is
\begin{align}
E_{i}^{CI}&=-\frac{2}{g}\frac{R^{2}}{(\rho^{2}+R^{2})^{2}}
\sigma_{i}\,,\;
E_{i}^{C\bar{I}}=\frac{2}{g}\frac{R^{2}}{(\rho^{2}+R^{2})^{2}}
\sigma_{i}\,,\\
B_{k}^{CI} & =B_{k}^{C\bar{I}}=-\frac{2}{g}
\frac{R^{2}}{(\rho^{2}+R^{2})^{2}}\sigma_{k}\,.
\end{align}
One can clearly see that the CYMI corresponds to ``parallel''
fields, whereas the anti-CYMI corresponds to ``anti-parallel''
fields in Minkowski spacetime (imaginary electric fields in Euclidean
spacetime). This property we will show gives rise to pair production.
Let us remark though that real Minkowski electric fields are not a 
sufficient requirement for pair production, a simple Abelian 
counterexample is provided through plane waves.
Before solving Wong's equations, let us discuss some basic
properties of the fields.

The CYMI are not self-dual, but one can show the totally
antisymmetric field strength tensor is related to an imaginary
anti-CYMI. To show this let us introduce `t Hooft symbols,
\begin{align}
\eta_{a\mu\nu}^C&=\varepsilon_{a\mu\nu4}
+i\delta_{a\mu}\delta_{\nu4}-i\delta_{a\nu}\delta_{4\mu}\,,\\
\bar{\eta}_{a\mu\nu}^{C}&=\varepsilon_{a\mu\nu4}
-i\delta_{a\mu}\delta_{\nu4}+i\delta_{a\nu}\delta_{4\mu}\,,
\end{align}
for $\sigma_{\mu\nu}^{C}=\frac{1}{2}\eta_{a\mu\nu}^{C}\sigma_{a}$
and $\bar{\sigma}_{\mu\nu}^{C}=\frac{1}{2}
\bar{\eta}_{a\mu\nu}^{C}\sigma_{a}$.
Then we can find the following identities:
\begin{align}
\widetilde{\eta}_{a\mu\nu}^{C}&
=\frac{1}{2}\varepsilon_{\mu\nu\alpha\beta}\eta_{a\alpha\beta}^{C}
=i\bar{\eta}_{a\mu\nu}^{C}\,,\\
\widetilde{\bar{\eta}}_{a\mu\nu}^{C}&
=\frac{1}{2}\varepsilon_{\mu\nu\alpha\beta}
\bar{\eta}_{a\alpha\beta}^{C}
=-i\eta_{a\mu\nu}^{C}\,.
\end{align}

To further explore the properties, let us introduce in matrix form
the `t Hooft symbols as combinatory rotation and translation 
(imaginary rotation)
generators of SO$(4)$. Namely we have that 
\begin{equation}
K_{a\mu\nu}= i\delta_{a\mu}\delta_{\nu4}
-i\delta_{a\nu}\delta_{4\mu},\quad 
L_{a\mu\nu} = \varepsilon_{a\mu\nu4}\,.
\label{eq:complex_gen}
\end{equation}
And we can write for the `t Hooft symbols
$\eta_{i}^{C}=L_{i}+K_{i}$ and $\bar{\eta}_{i}^{C}=L_{i}-K_{i}$.
We have that $K_{i}^{T}=-K_{i}$ and $L_{i}^{T}=-L_{i}$. The 
generators satisfy the following algebra:
\begin{align}
[K_{i},K_{j}]&=\varepsilon_{ijk}L_{k}\,,\;
[L_{i},L_{j}]=-\varepsilon_{ijk}L_{k}\,,\\
\quad[L_{i},K_{j}]&=-\varepsilon_{ijk}K_{k}\,.
\end{align}
One may also show by introducing the tensor,
$\Delta_{\mu\nu}\coloneqq \delta_{\mu\nu}
-2\delta_{\mu 4}\delta_{\nu 4}$,
the following relation holds:
$\{L_{i},K_{j}\}=\varepsilon_{ijk}K_{k}\Delta$.
Note, we also have that $\{L_{i},L_{j}\}=\{K_{i},K_{j}\}$ for $i\neq j$.
Last, one can determine that
\begin{equation}
\{\eta_{i}^{C},\bar{\eta}_{j}^{C}\}=
2\varepsilon_{ijk}\Delta K_{k}-2\delta_{ij}\,.
\label{eq:etaetabar}
\end{equation}

Using the generators, Eq.~\eqref{eq:complex_gen}, one may 
directly find that not only do the CYMI have finite energies, but also
that they vanish,
\begin{equation}
\textrm{tr}G_{\mu\nu}^{CI}G_{\nu\mu}^{CI}=0\,,\quad 
\textrm{tr}G_{\mu\nu}^{C\bar{I}}G_{\nu\mu}^{C\bar{I}}=0\,.
\end{equation}
One may also, using Eq.~\eqref{eq:etaetabar}, confirm that we have
\begin{equation}
\textrm{tr}[G_{\mu\nu}^{CI}\widetilde{G}_{\nu\mu}^{CI}] 
 =-\frac{4\cdot24i}{g^{2}}\frac{R^{4}}{(\rho^{2}+R^{2})^{4}}\,,
\end{equation}
and then in a Euclidean picture,
\begin{equation}
\frac{g^{2}}{16\pi^{2}}\int d^{4}x\textrm{tr}G_{\mu\nu}^{CI}
\widetilde{G}_{\mu\nu}^{CI} =-1\,,
\end{equation}
in agreement with the Minowski definition, 
Eq.~\eqref{eq:real_time_winding}. Likewise we have that
$\frac{g^{2}}{16\pi^{2}}\int d^{4}x\textrm{tr}G_{\mu\nu}^{C\bar{I}}
\widetilde{G}_{\mu\nu}^{C\bar{I}} =1$.

The CYMI, however, does not solve the Yang Mills equation of motion. In
fact one can find using the above identities that
\begin{equation}
[\mathcal{D}_{\mu}^{CI},G_{\mu\nu}^{CI}] =\frac{8}{g}
\frac{R^{2}}{(\rho^{2}+R^{2})^{3}}
\bigl\{ x^{T}\Delta(L+K)\cdot\sigma-x^{T}K\cdot\sigma\bigr\}_\nu\,.
\end{equation}
However the Bianchi identity holds as it should, 
$[\mathcal{D}_{\mu}^{CI},\widetilde{G}_{\mu\nu}^{CI}]=0$.

Wong's equations in a CYMI written in a Lorentz index matrix 
representation read,
\begin{align}
\dot{I}^{CI}_c&=-\frac{2}{\rho^{2}+R^{2}}
\varepsilon_{abc}\dot{x}^{CI\,T}\eta_{a}^{C}x^{CI}I^{CI}_{b}\,,
\label{eq:isospin_CYMI}\\
\ddot{x}^{CI}&=\frac{i|\dot{x}^{CI}|}{m}\frac{2R^{2}}{(\rho^2+R^{2})^{2}}
\eta^{C}\cdot I^{CI}\dot{x}^{CI}\,,
\end{align}
and we can evaluate them similar to as was accomplished in 
Sec.~\ref{sec:aa_YMI}, namely through the usage of a large parameter
limit for the CYMI coupled with the adiabatic theorem
in isospin.

\subsection{Adiabatic Theorem and the Large Complex Instanton}
\label{sec:sec:aa_CYMI}

Employing the adiabatic theorem one can determine a gauge
element, $u^{CI}\in$ SL$(2,\mathbb{C})$, such that the isospin equation 
of motion, Eq.~\eqref{eq:isospin_CYMI}, is satisfied and
\begin{equation}
u^{CI}A^{CI}\dot{x}^{CI}(u^{CI})^{-1}=
\sqrt{A^{CI\,a}\dot{x}^{CI}A^{CI\,a}\dot{x}^{CI}}\sigma_3\,;
\end{equation}
c.f., Eq.~\eqref{eq:aa_gauge}. This gives us
\begin{equation}
I^{CI}_a=\frac{\dot{x}^{CI\,T}\eta_{a}^Cx^{CI}}
{\sqrt{\dot{x}^{CI\,T}\eta^C_{a}x^{CI}\dot{x}^{CI\,T}\eta_{a}^Cx^{CI}}}\,.
\end{equation}
The isospin for the CYMI is an element of the coset 
SL$(2,\mathbb{C})/$SU$(2)$. Again we have propertime independence,
$\dot{I}^{CI}=0$. And that we may apply the adiabatic theorem and in 
turn ignore off-diagonal parts to Berry's phase, owing to the 
antisymmetric tensors, $\eta_A^C$, one must have that 
$x^{CI}\propto \ddot{x}^{CI}$.

For the evaluation of the Lorentz force equation we examine the 
large complex instanton limit such that $R^2\gg \rho^2$. And we 
have that
\begin{equation}
\ddot{x}^{CI}\approx\frac{2i|\dot{x}^{CI}|}{mR^2}
\eta^{C}\cdot I^{CI}\dot{x}^{CI}\,,
\label{eq:cymi_lorentz_large}
\end{equation}
which we evaluate using electric and magnetic projection operators:
\begin{equation}
P_{B}^{CI}\coloneqq \frac{1}{2}[1-(\eta^{C}\cdot I^{CI})^{2}]\,,\quad 
P_{E}^{CI}\coloneqq \frac{1}{2}[1+(\eta^{C}\cdot I^{CI})^{2}]\,,
\end{equation}
such that,
\begin{equation}
(\eta^{C}\cdot I^{CI})^{2}P_{B}^{CI}=-P_{B}^{CI}\,,\quad
(\eta^{C}\cdot I^{CI})^{2}P_{E}^{CI}=P_{E}^{CI}\,.
\end{equation}
with the properties of idempotency, $(P_{B,E}^{CI})^{2}=P_{B,E}^{CI}$, 
and also that $P_{E}^{CI}+P_{B}^{CI}=1$ and $P_{E}^{CI}P_{B}^{CI}=0$.
We may use the projection operators to decouple the Lorentz force
equation, and moreover the coordinates into electric and magnetic 
degrees of freedom,
\begin{equation}
x_E^{CI}=P_E^{CI}x^{CI}\,,\quad x_B^{CI}=P_B^{CI}x^{CI}\,.
\end{equation}
The Lorentz force equation may be readily solved using projection
operators; see~\cite{Fradkin_1978} for details. 
Alternatively, one may simply find
an SO$(4)$ transformation such that the coordinates in 
Eq.~\eqref{eq:cymi_lorentz_large} are parallel in an arbitrary direction,
which also projects the electric and magnetic parts, and leads to the 
exact same result.
Solutions of the decoupled Lorentz force yield
\begin{align}
\dot{x}^{CI}_E(\tau)&=
\Bigl[\cos\bigl(\frac{2|\dot{x}^{CI}|}{mR^{2}}\tau\bigr)
+i\eta^{C}\cdot I^{CI} 
\sin\bigl(\frac{2|\dot{x}^{CI}|}{mR^{2}}\tau\bigr)\Bigr]
\dot{x}_{E}^{CI}(0) \label{eq:cymi_x_dot} \\
\dot{x}^{CI}_B(\tau)&=\Bigl[\cosh\bigl(
\frac{2|\dot{x}^{CI}|}{mR^{2}}\tau\bigr)
\nonumber \\ &\quad+i\eta^{C}\cdot I^{CI} 
\sinh\bigl(\frac{2|\dot{x}^{CI}|}{mR^{2}}\tau\bigr)\Bigr] 
\dot{x}_{B}^{CI}(0)\,.
\end{align}
The situation here is analogous to the one encountered for homogeneous
fields in Sec.~\ref{sec:homo}. Namely, we cannot both satisfy both
equations given the periodic boundary conditions; 
to maintain a real stationary 
point and hence real $|\dot{x}^{CI}|$ we take the magnetic components
trivial, i.e., $x_B^{CI}=\text{constant}$. The electric components, given 
the circular instanton constraint through $x^{CI}\propto \ddot{x}^{CI}$, 
can be found as
\begin{equation}
x_E^{CI}(\tau)=\frac{imR^{2}}{2|\dot{x}^{CI}|}\bar{\eta}^{C}\cdot I^{CI}
\dot{x}_E^{CI}(\tau)\,.
\label{eq:cymi_x}
\end{equation}
Then the WIs are solely determined
from the electric fields. We can immediately write down the WI
periodic criteria as,
\begin{equation}
|\dot{x}^{CI}|=n\pi mR^{2} \quad \forall n\in \mathcal{Z}^+\,, 
\label{eq:cymi_schwinger}
\end{equation}
as anticipated, c.f., Eq.~\eqref{eq:cymi_criteria}. 
We do have pair production for the 
CYMI, the topological fields decay by a vacuum instability. 

The Schwinger effect exponential suppression is 
$(1/2)n\pi m^2R^{2}$, 
understood from the worldline action, Eq.~\eqref{eq:S_n}. 
One can calculate using
Eq.~\eqref{eq:cymi_x_dot} and Eq.~\eqref{eq:cymi_x} the contribution 
to the worldline action of
$\int_{0}^{1}d\tau\,\textrm{tr}\bigl[I^{CI}igA^{CI}\dot{x}^{CI}\bigr]
=(m/2)|\dot{x}^{CI}|$. The Berry's phase term, here too, only 
introduces a trivial factor of $4\pi i$.

\section{Conclusions}
\label{sec:conclusions}

Schwinger pair production has been analyzed in the topological
BPST instanton (YMI), and due to the Hermiticity of its 
construction, no vacuum decay via the Schwinger effect was
found as anticipated. However, as an anomaly cancellation is
found for Abelian homogeneous fields, (one which is 
revived through the Schwinger mechanism), it is likewise
anticipated that a non-Abelian field configuration with 
Chern-Pontryagin density should be present and decay via the 
Schwinger effect. We construct such a field that is gauge 
invariant in SL$(2,\mathbb{C})$, in a Euclidean metric, and
SU$(2)$, in a Minkowski metric. The field resembles parallel
or anti-parallel fields in Minkowski spacetime. To accomplish 
calculations, we extended the WI method to non-Abelian fields.

The WI method is important for the study of Schwinger pair
production in inhomogeneous fields. Thus, (apart from 
the study of the Schwinger effect in YMI/CYMI), our two-fold 
scope included the development of the WI method for a
generically complex SL$(2,\mathbb{C})$ background field. 
To arrive at Wong's equations--the non-Abelian equivalent of
the Lorentz force equation--we made use of the coherent 
state method. There, color degrees of freedom were summed over 
in a Haar measure extended for the non-compact group.

The WI method in non-Abelian systems may prove useful for not only the 
fields discussed in this work, but also for color-glass 
condensate~\cite{doi:10.1146/annurev.nucl.010909.083629}
backgrounds, as are thought present in the early stages of heavy-ion
collisions. Furthermore, we expect the non-Abelian WI method to be 
essential in the development of the worldline formalism to handle a variety 
of topological field theories through the construction of bulk or boundary
$G/H$ type coset theories, such as the Wess-Zumino-Witten, conformal 
sigma models, and Chern-Simons
theories~\cite{Gawedzki:2001ye,Gawedzki:1991yu}.
Last, a realization of large $N$ SU$(N)$ Yang Mills 
and nonlinear sigma models 
through coherent states has been reported in~\cite{10.1143/PTP.104.1189},  
presenting us with an array of chiral models which would be worthwhile to 
revisit under the light of the WI method introduced here.

\appendix*
\section{Exact Non-Persistence in
 Abelian-like Homogeneous Parallel Fields}
\label{sec:appendix}

Here we evaluate the contributions to Schwinger pair production 
coming from the effective action exactly in non-Abelian homogeneous
parallel fields given by Eq.~\eqref{eq:homo_G}. 
This is accomplished by summing over the eigenvalues 
of the worldline Hamiltonian. Let us begin by writing the effective 
action, Eq.~\eqref{eq:eff_action_def}, in Schwinger proper time as
$\Gamma[A]=\tr\int_0^\infty \frac{dT}{T}\exp[-(-D^{2}+m^{2})T]$.
The homogeneous fields are given in Eq.~\eqref{eq:homo_fields}.
It proves useful to first take the color trace, summing over both colors.
This is permissible as the Hamiltonian
is already diagonal in color. The effective action becomes
\begin{align}
\Gamma[A]&=\int_{0}^{\infty}\frac{dT}{T}e^{-m^2T}
\int d^4x\langle x |[e^{D_{+}^{2}T} + e^{D_{-}^{2}T}]| x \rangle\,,\\
D_\pm^2&=\partial_{1}^{2}+\partial_{3}^{2}+(\partial_{2} \mp 
igBx_{1})^{2}+(\partial_{4} \mp i^{2}gEx_{3})^{2}.
\label{eq:Dpm}
\end{align}

Let us define a Euclidean
Fourier transform such that 
$p\langle x | p\rangle = -i\partial / \partial x\langle x | p \rangle$,
$\langle x_2 | p_2 \rangle \sim \exp(ip_2x_2)$, 
$\int\frac{dp}{2\pi}|p\rangle\langle p |=1$,
and $\langle p | p' \rangle =2\pi \delta(p-p')$. 
Upon insertion of complete sets of states, 
one may then find the eigenvalues of Eq.~\eqref{eq:Dpm} as
$E_{\pm\,n,m}=\mp 2\bigl(n+\frac{1}{2}\bigr)gB 
\mp 2\bigl(m+\frac{1}{2}\bigr)igE$,
for $n,m\in[0,\infty)$. Upon summing over the eigenvalues
one finds for the effective action
\begin{equation}
\Gamma[A]=\frac{i}{2}\int_0^{\infty}\frac{dT}{T}
\int dx_{2,4}\int\frac{dp_{2,4}}{(2\pi)^{2}}
\frac{\exp(-m^2T)}{\sinh(gBT)\sin(gET)}\,,
\end{equation}
where the contribution from $D_{-}^{2}$ is the same as that
of $D_{+}^{2}$. The coordinate and momenta integrals are divergent;
this is due to our selection of homogeneous fields which are 
un-bounded.
We may normalize the action by considering a closed box with Landau
modes such that $\int dx_{2}=L$ and $\int dp_{2}=gBL$, and likewise
for Euclidean time for general length $L$. Hence
\begin{equation}
\Gamma[A]=\frac{2ig^2EBL^{4}}{(4\pi)^2}\int_{0}^{\infty}
\frac{dT}{T}\frac{\exp(-m^2T)}{\sinh(gBT)\sin(gET)}\,.
\end{equation}
We can find the contribution which pertains to Schwinger pair 
production by noting that $\textrm{Im} \Gamma_{Minkowski}[A]
=\textrm{Re} \Gamma_{Euclidean}[A]$. 
This can be found through the poles on the real positive axis in the sine
function. Taking the residues one can find the exact to one-loop pair 
production non-persistance as 
\begin{equation}
\textrm{Re}\Gamma[A]=\frac{g^{2}EBL^{4}}{8\pi^2}
\sum\limits _{n=1}^{\infty}\frac{(-1)^{n+1}}{n}
e^{-\frac{\pi nm^{2}}{gE}}\sinh^{-1}
\bigl(\frac{n\pi B}{E}\bigr)\,.
\label{eq:homo_exact}
\end{equation}

\begin{acknowledgments}
We would like to thank James P. Edwards, Yoshimasa Hidaka, and 
Shi Pu for valuable discussions. We are also grateful to the University of 
Tokyo where a portion of this work was accomplished. P.C. would also 
like to thank the Theory Center of KEK where a portion of this work was 
accomplished and Fudan University for facilitating a portion of this work.
\end{acknowledgments}

\bibliography{references}
\bibliographystyle{apsrev4-1}
\end{document}